\documentclass[traditabstract]{aa} 
\usepackage{latexsym}
\usepackage{amssymb}
\usepackage{lscape}
\usepackage[]{natbib}

\usepackage{color}

\usepackage{graphicx}
\usepackage{txfonts}
\def\lsim{\mathrel{\rlap{\lower 3pt \hbox{$\sim$}} \raise 2.0pt \hbox{$<$}}}
\def\gsim{\mathrel{\rlap{\lower 3pt \hbox{$\sim$}} \raise 2.0pt \hbox{$>$}}}
\title{Ubiquitous ram pressure stripping in the Coma cluster of galaxies\thanks{based on observations taken at the San Pedro Martir 
telescope belonging to the Mexican National Observatory (OAN)}}
\subtitle{}
\author{G. Gavazzi \inst{1} 
\and G. Consolandi \inst{1}                                                                
\and M. L. Gutierrez \inst{2}                            
\and A. Boselli \inst{3}                                    
\and M. Yoshida \inst{4}
}
\authorrunning{G. Gavazzi  et al.}
\titlerunning{Three new H$\alpha$ tails in the Coma cluster}
\institute{Universit\`a degli Studi di Milano-Bicocca, Piazza della Scienza 3, 20126 Milano, Italy\\
\email {giuseppe.gavazzi@mib.infn.it}
\and
nstituto de Astronomia, UNAM, Km 107 Carretera Tijuana-Ensenada, Ensenada, B.C., Mexico 22860
\and
Aix Marseille Univ, CNRS, CNES LAM, Laboratoire d'Astrophysique de Marseille, Marseille, France
\and
Subaru Telescope, National Astronomical Observatory of Japan,
National Institutes of Natural Sciences,
650 A'ohoku Place, Hilo, Hawaii 96720, USA
}
\begin{document}

\date{Received; accepted}
\abstract{We report the detection of H$\alpha$ trails behind three new intermediate-mass irregular galaxies  in the NW outskirts of the nearby cluster of galaxies
Abell 1656 (Coma). Hints that these galaxies possess an extended component were found in earlier, deeper H$\alpha$ observations carried out with the Subaru telescope.
However the lack of a simultaneous $r$-band exposure, together with the presence of strong stellar ghosts in the Subaru images, prevented us from 
quantifying the detections. We therefore devoted one full night of H$\alpha$ observation to each of the three galaxies using the San Pedro Martir 
2.1m telescope. 
One-sided tails of H$\alpha$  emission of  10-20 kpc projected size were detected, suggesting  an ongoing ram pressure stripping event. We added these 3 new sources 
of extended ionized gas (EIG) added to the 12 found by Yagi et al. (2010), NGC 4848 (Fossati et al. 2012), 
and NGC 4921 whose ram pressure stripping is certified by HI asymmetry. This  brings the number sources with  H$\alpha$ trails to  
17 gaseous tails out of 27 (63 \%) late-type galaxies (LTG) galaxies members of the Coma cluster
with direct evidence of ram pressure stripping. The 27 LTG galaxies, among these the 17 with extended H$\alpha$ tails, have kinematic properties  that are different 
from the rest of the early-type galaxy (ETG) population of the c ore of the Coma cluster, as they deviate  in the phase-space diagram $\Delta$V/$\sigma$ versus $r/R_{200}$.} 
\maketitle

\section{Introduction}

A 1.5 $\rm deg^2$ region of the nearby Coma cluster of galaxies (Abell 1656, $<z>\sim0.023$, $<cz>\sim 6900$ $\rm km~s^{-1}$ , Dist=100 Mpc) was surveyed with deep
H$\alpha$ observations using the Subaru telescope (Yagi et al. 2010,   see the inset of Figure \ref{plot} showing the footprint of the 
Subaru field).
These observations, carried out at the limiting surface brightness of $2.5\times 10^{-18}~\rm erg ~cm^{-2}~sec^{-1}~arcsec^{-2}$,
revealed the presence of H$\alpha$ extended ionized gas (EIG) behind 14 galaxies. 
This indicates that, when observed with sufficiently sensitive exposures, 
approximately 50 \% of all LTGs in this cluster reveal an associated extended trail of H$\alpha$, 

The field of Yagi et al. (2010), however, did not cover westward of  RA $12^h58^m25^s$. For example the bright galaxy NGC 4848 was not covered by these 
observations and a special five hour pointing with the San Pedro Martir (SPM) 2.1m telescope by Fossati et al. (2012) revealed the presence of an H$\alpha$
trailing emission of 65 kpc projected length.
Similarly it did not cover the cluster  eastward of RA $13^h01^m00^s$. The bright galaxy NGC 4921 was not included, but by combining
Hubble Space Telescope (HST) imaging with HI line mapping, Kenney et al. (2015) found a significant displacement of the neutral hydrogen from stars; this displacement is indicative of ram pressure stripping.
Another H$\alpha$ Subaru pointing of the NW region of the Coma cluster was obtained in 2014 (Yagi, private communication),
however the absence of a corresponding broadband ($r$) exposure to estimate and subtract the continuum emission, prevented a proper
analysis of this field.
However this observation hinted at possible unilateral  H$\alpha$ emission behind 3 additional galaxies J125750.2+281013, J125756.7+275930,  and  J125757.7+280342
in the northwestern part of the cluster.
Inspired by these tentative detections we decided to devote approximately one night of observation at the SPM observatory
to each of these galaxies in April, 2018. 
We report the positive detection of extended, asymmetric H$\alpha$ emission associated with each of these 
\footnote{Note added in proofs: the three galaxies were reported as possible candidates for ram pressure stripping 
by Smith et al. (2010), based on GALEX selection.}.
In total the core region of the Coma cluster covered with H$\alpha$ observations contains 203 early-type galaxies (E+S0+S0a; ETGs), 
27 late-type galaxies (LTGs), 17 of which show evidence of extended H$\alpha$. 
Assuming  that ionized tails arise from ram pressure stripping (Gunn \& Gott, 1972) 
of galaxies crossing the intracluster medium (ICM) at high speed for the first time, and that the gas ablation produced by such interaction
proceeds on timescales as short as 100 Myr (Boselli \& Gavazzi, 2006, 2014), we confirm that 
a massive infall of gas-rich, star forming galaxies 
(100-400 galaxies per Gyr, Adami et al. 2005, Boselli et al. 2008; Gavazzi et al. 2013, 2013b, 2017) 
is currently occurring onto rich clusters of galaxies such as Coma.

\begin{figure}
\centering
 \includegraphics[width=8.5 cm]{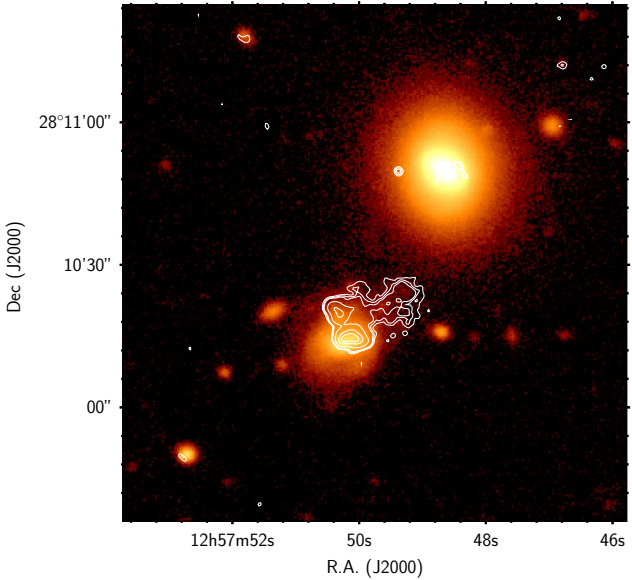}
\caption{Grayscale representation of the $r$-band emission from J125750.2+281013. Contours represent the H$\alpha$ NET emission. The proximity of the massive galaxy 
CGCG 160-049 does not allow us to exclude that the gaseous asymmetry is due to tidal interaction. However we note that stars appear not accordingly displaced. 
Levels are -17.07 -16.89 -16.29 -15.87 -15.69 $\rm erg~cm^{-2} sec^{-1} arcsec^{-2}$ in log units, after three pixel Gaussian smoothing.}
\label{tail1} 
\end{figure}

\begin{figure}
\centering
\includegraphics[width=8.5 cm]{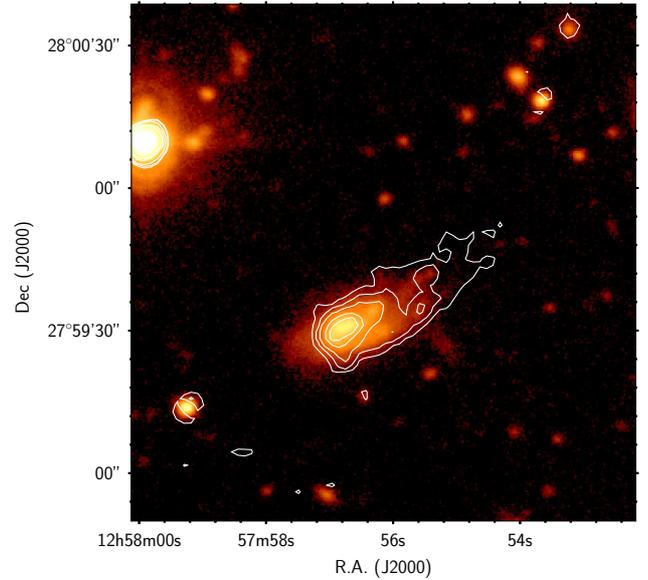}
\caption{Same as Figure \ref{tail1} for  J125756.7+275930. 
Levels are  -17.30 -17.07 -16.77 -15.77 -15.47  $\rm erg~cm^{-2} sec^{-1} arcsec^{-2}$ in log units, after three pixel Gaussian smoothing.}
\label{tail3}  
\end{figure}

\begin{figure}
\centering
\includegraphics[width=8.5 cm]{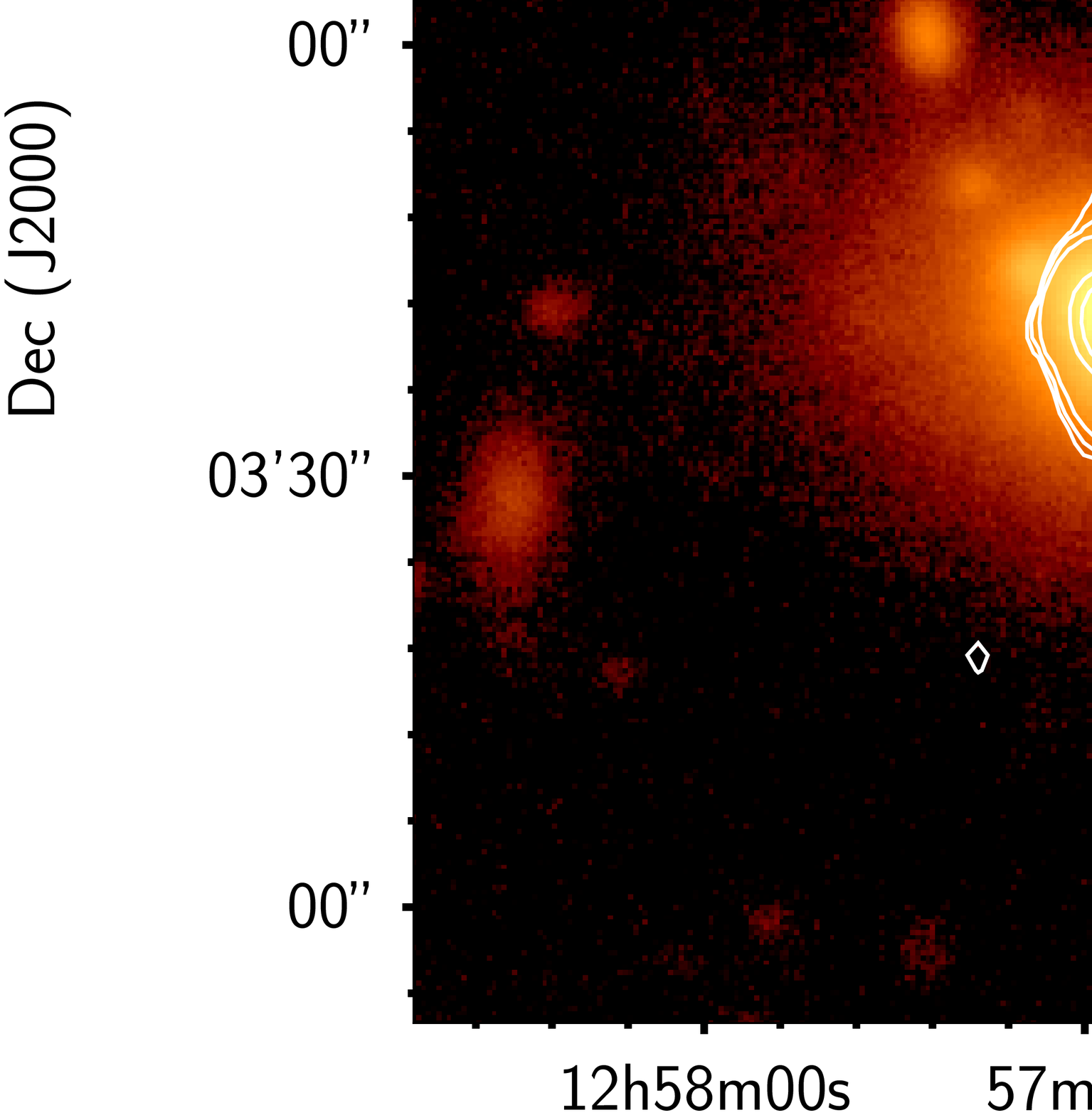}
\caption{Same as Figure \ref{tail1} for  J125757.7+280342.
Levels are -17.30 -16.77 -16.30 -15.77 -15.47 $\rm erg~cm^{-2} sec^{-1} arcsec^{-2}$ in log units, after three pixel Gaussian smoothing.}
\label{tail2}  
\end{figure}

\begin{table*}
\caption{Observational parameters of the  target galaxies.}                                                                   
\centering                                                                                        
\begin{tabular}{l c c c c c c c c}                                                                    
\hline\hline                                                                                      
  Object  &    RA                &    Dec       &    $cz$           & Log$ M_*$      &  $g-i$ &  H$\alpha$   filter & Texp ON  &  Texp OFF       \\                            
          &  (hh m s)            & ($^o$ ' ")   & $ \rm km~s^{-1}$       & $M\odot$  & mag  &    \AA              &  min     &   min            \\                           
\hline                                                                                            
 J125750.2+281013   &    12 57 50.2       &   28 10 13  &  6936  & 8.47 &  0.22    &  6723  & 360 & 30    \\
 J125756.7+275930   &    12 57 56.7       &   27 59 30  &  4579  & 8.53 &  0.22    &  6683  & 360 & 30    \\
 J125757.7+280342   &    12 57 57.7       &   28 03 42  &  8147  & 9.75 &  0.78    &  6723  & 480 & 40    \\
\hline 
\end{tabular}                                                                                     
\label{tab1}                                                                                      
\end{table*}    

\begin{table*}
\caption{Derived  H$\alpha$+[NII]  parameters of the  target galaxies in the 3 arcsec aperture.}                                                                   
\centering                                                                                        
\begin{tabular}{l c c c c c c}                                                                    
\hline\hline                                                                                      
  Object            &    EW 3'' &   $\pm$  &   Log   Flux 3''           &   $\pm$                       &   EW SDSS  &   Log   flux  SDSS            \\                
                    &  \AA    &   \AA & $ \rm erg~cm^{-2} sec^{-1}$ & $ \rm erg~cm^{-2} sec^{-1}$  &   \AA    &   $ \rm erg~cm^{-2} sec^{-1}$  \\      
\hline                                                                                            
 J125750.2+281013   &   -25.8   &   3.4   &       -14.75                &      0.05             &  -36.6     & -14.68                        \\
 J125756.7+275930   &   -90.4   &   4.7   &       -14.25                &      0.02                     &  -100.8    & -14.24                        \\
 J125757.7+280342   &   -63.2   &   4.2   &       -14.04                &      0.02             &  -45.9     & -14.14                        \\
\hline 
\end{tabular}                                                                                     
\label{tab2}                                                                                      
\end{table*}    

\begin{table*}
\caption{Derived total  H$\alpha$+[NII]  parameters of the  target galaxies.}                                                                   
\centering                                                                                        
\begin{tabular}{l c c c c c c}                                                                    
\hline\hline                                                                                      
  Object            &  EW tot  &   $\pm$ &   Log   flux tot            &   $\pm$   &  Limiting $\Sigma$  & Proj. length\\                    
                    &  \AA   &   \AA & $ \rm erg~cm^{-2} sec^{-1}$  &   $ \rm erg~cm^{-2} sec^{-1}$ & $ \rm erg~cm^{-2} sec^{-1} arcsec^{-2}$  & kpc     \\       
\hline                                     
 J125750.2+281013   &   -13.7  & 3.1     &-14.01                       &  0.09     &   5.88 $\times 10^{-18}$   &  10   \\       
 J125756.7+275930   &   -56.3  & 3.7     &-13.47                       &  0.02     &   5.88 $\times 10^{-18}$   &  22   \\       
 J125757.7+280342   &   -20.3  & 3.4     &-13.28                       &  0.07     &   5.01 $\times 10^{-18}$   &  24   \\       
\hline 
\end{tabular}                                                                                     
\label{tab3}                                                                                      
\end{table*}

\section{Observations}
\label{Obs}
On the nights of April 12, 13, and 14, 2018, we used the 2.1 m telescope at SPM to repeatedly observe  three 5 x 5 $\rm arcmin^2$ regions of the
nearby Coma cluster of galaxies, each targeting one of the J125750.2+281013, J125756.7+275930, J125757.7+280342  galaxies.
 We used two narrowband (80 \AA) filters centered at $\lambda$ 6683 and $\lambda$ 6723 \AA ~to detect the H$\alpha$ emission (at the redshift  listed 
 in Table \ref{tab1}
 along with other parameters taken from GOLDmine; Gavazzi et al. 2003)  
and a broadband $r$ (Gunn) filter (effective $\lambda$ 6231 \AA, $\Delta\lambda \sim 1200$ \AA) to recover the continuum emission.

The fields were observed with several individual pointings of 1800 seconds in the ON-band filter,
and with same number of 300 seconds exposures with the broadband (OFF-band) $r$ filter, up to a total exposure time given in Table \ref{tab1}. 
After flat fielding, the aligned observations were combined 
into a final ON-band frame and an OFF-band frame (details on the reduction procedures of H$\alpha$ observations can be found in Gavazzi et al. 2012).

To calibrate the data, we repeatedly observed the spectrophotometric stars Feige34 and Hz44 (Massey et al. 1988). From the calibrated data we extracted
the flux and the EW within a circular aperture of 3 arcsec (diameter) and we checked the calibration comparing these values with
the  measurements of H$\alpha$+[NII] flux and EW  extracted from the  Sloan Digital Sky Survey  (SDSS) nuclear fiber spectra (see Table \ref{tab2}).
In spite of the uncertainty in the absolute positioning of the central aperture, 
the agreement between two sets of data is satisfactory, reassuring us concerning the quality of the photometric calibration.
The integrated photometric parameters are  listed in Table \ref{tab3}.
Among these the projected length gives the approximate full extension of the H$\alpha$ tail from the center of the parent galaxy.

In Table \ref{tab1} we list the celestial coordinates, redshift, total stellar mass, 
derived from the $i$-band luminosity and the $g-i$ color according to Zibetti et al. (2009), assuming a Chabrier initial mass function (IMF) (Chabrier 2003).
Our observations, owing to the generous integration time of 6/8 hours (ON band), reached a deep sensitivity limit and provided the estimate of 
other extended parameters, as listed in Table \ref{tab3}.\\

\section{Estimate of ionized gas mass}

An estimate of the density of the ionized gas in the tail of J125756.7+275930, the most clear cut candidate for ram pressure stripping,  
can be achieved using equation (1) and (2) in Boselli et al. (2016a).
The stripped material is assumed to be distributed in a cylinder of diameter of 5 kpc and of length comparable to the deprojected extension of the 
tail of ionized gas.   
Since the galaxy has a redshift of 4579 $\rm km~s^{-1}$ ( as compared  to the mean cluster velocity of 6900 $\rm km~s^{-1}$), it  
is crossing the cluster from behind. Thus the real length of the tail can be roughly estimated multiplying the observed length
by $\sqrt 3$, obtaining 38 kpc.
The region containing the stripped gas (assumed with a filling factor of 0.1) has a volume of 710 $\rm kpc^3$ and a luminosity 
$L({H\alpha})$ of  $2.44x10^{40}$ $\rm erg sec^{-1}$, assuming [NII]/H$\alpha$=0.1 and 70\% of the total H$\alpha$ emission in the tail. 
Thus the region has a density of electrons (or protons) of  0.19 $\rm cm^{-3}$, which translates into
a total mass of $3.05*10^8$ $M_{\odot}$ of ionized gas. This is consistent with $3.55*10^8$ $M_{\odot}$ MHI lost in the tail, roughly estimated 
from the measured $\rm MHI=10^{8.62}$ $M_{\odot}$, combined with a deficiency parameter of $\rm Def_{HI}=0.27$.

\section{Discussion}

The full H$\alpha$ survey of the central region of the Coma cluster (covering approximately
$12^h57^m-13^h02^m; +27^d30^p-28^d17^p$),  including the work of Yagi et al. (2010), was extended to the NW by Fossati et al. (2012) and by the present work.
Assuming the $r$=17.7 mag threshold adopted by SDSS, this sky window contains   203 ETGs (E/S0/S0a) and 27 LTGs (Sa-Sm-BCD) (as listed in Table \ref{tabLTG}, 
taken from GOLDMine).
Among the LTGs, 12 are found associated with cometary H$\alpha$ tails from Yagi et al. (2010) \footnote{The full number of H$\alpha$ tails
in Yagi et al. (2010) is 14, but two are associated with ETGs, as listed in the last lines of Table \ref{tabLTG}}. One (N4848) was found extended 
by Fossati et al. (2012), and 3 in this work. In addition,  NGC 4921, while not showing extended 
H$\alpha$ emission, was considered an example of galaxy subject to ram pressure stripping because of its asymmetric 
HI distribution by Kenney et al. (2015). In total there are 17/27 (63\%) LTGs (plus 2 ETGs) with H$\alpha$  cometary tails or HI asymmetry 
indicating ram pressure. In conclusion more than one out of two LTGs, possibly gas rich systems in this cluster, show  cometary H$\alpha$ debris, 
which is a signature of ram pressure stripping.
This reinforces a previous result for A1367 where 11/26 or 42\% of the surveyed LTGs show H$\alpha$ tails (Consolandi et al. 2018).
 \\
Some properties of the 27 surveyed LTG galaxies (+ 2 ETG with tails) are listed in Table \ref{tabLTG}. In particular we
list the $r$ magnitude and the $g-i$ color (SDSS, as given by GOLDMine) and the HI deficiency parameter.
This parameter was defined by Haynes \& Giovanelli (1984) as $\rm Def_{HI}=<logMHI(T,dL)> - logMHI(obs)$
where <logMHI(T,dL)>=C1+C2x2 log(dL), where dL [kpc] is determined in the g-band at the $\rm 25^{th} ~mag ~arcsec^{-2}$ isophote.
References to the HI measurements can be found in Gavazzi et al. (2006), however the deficiency parameter is recomputed in this work adopting C1=7.51 and C2=0.68 for all LTGs (Sa-BCD) (as discussed by Gavazzi et al. 2013).
The HI deficiency parameter is unfortunately available only for 10/27 galaxies. It is noticeable that all available values are found between 0.25 and 1.01, implying an HI deficiency factor between 1.8 and 10 times the normal HI content of LTGs.  

The spatial distribution and pointing direction of the H$\alpha$ tails detected by Yagi et al. (2010) was sketched in their work (see their Figure 6).
To further stress  the contribution of the present work 
a smaller area of the cluster is sketched in Figure \ref{coma} where the distribution of galaxies, 
as derived from the SDSS is given in grayscale.
Superposed are the X-ray contours of the hot gas emission (as revealed by XMM) embedding the galaxies. 
The head-tail radio galaxy 5C 04.081 is represented in red contours.
Blue contours give the H$\alpha$ extended emission from either Yagi et al. (2010), Fossati el al. (2012), or
the present work.
All tails lying in the NW part of the cluster, including NGC 4848, point toward NW, indicating  that the prevailing infall direction
is  from NW to SE.
The third dimension (along the line of sight) is however unclear: out of the four galaxies in this part of the cluster,
two have almost no velocity difference with respect to the cluster (suggesting motion in the plane of the sky), the other two
have opposite relative line-of sight velocities (-2321, +1247 $\rm km~s^{-1}$), which do not indicate a common provenance.

We quantify in Fig. 4 the kinematical difference of the virialized galaxies (ETGs) from all LTGs (probably  infalling 
into the cluster) and from LTGs that are currently 
losing their gas from ram pressure, as shown by the appearance of unilateral
gaseous tails. This is shown with the phase-space test of Boselli et al. (2014) and Jaff{\'e} et al. (2015) on the core region of 
the Coma cluster surveyed in 
H$\alpha$. Considering all SDSS galaxies
with $r$ brighter than  17.7 mag, this region contains  203 ETGs (E/S0/S0a, of which two with extended H$\alpha$  tails), 27 LTGs, of which 17 display 
an H$\alpha$ cometary tail.
It is self evident that LTGs in general, and in particular tailed LTGs, differ  from the distribution of 
ETGs, i.e., at any radius the 
LTGs (and even more the tailed LTGs) 
show higher peculiar normalized velocities, closer to the cluster escape velocity.
Disregarding the radial dependence of |$\Delta$V|/$\sigma$, we derive overall
$<|\Delta$V|/$\sigma>$=0.94 for ETGs, 1.53  and 1.28  for LTGs  with and without H$\alpha$ tails, respectively.
We must stress, however, that at large $ r/R_{200} \sim 0.4$, where two new NW LTG tails and NGC 4848 are found, their normalized velocity 
is $\leq 0.4$, which is much below the escape velocity; this is because  they were not in infall, or their infall was in the plane of the sky.
Finally, we note that the two ETGs, albeit empty from star formation at their interior, show extended ionized gas (see the last two lines of Table  \ref{tabLTG}).
It is possible that their gas was stripped in a previous (preprocessing) phase.
We conclude by speculating that our analysis is consistent with the idea that all LTGs are currently infalling, but 40\% of them are not seen with extended cometary tails
because they have not actually entered the dense intergalactic medium (IGM) of the cluster, in spite of being projected on it, or because they have been totally stripped in the past. 
The conspicuous infall rate that we observe in Coma and A1367 give us hope that a similar, or even higher rate exists in the Virgo cluster,
and will be revealed by the foreseen Virgo Environmental Survey
Tracing Ionised Gas Emission (VESTIGE)  announced by Boselli et al. (2018).

\begin{figure}
\centering
 \includegraphics[width=8.5 cm]{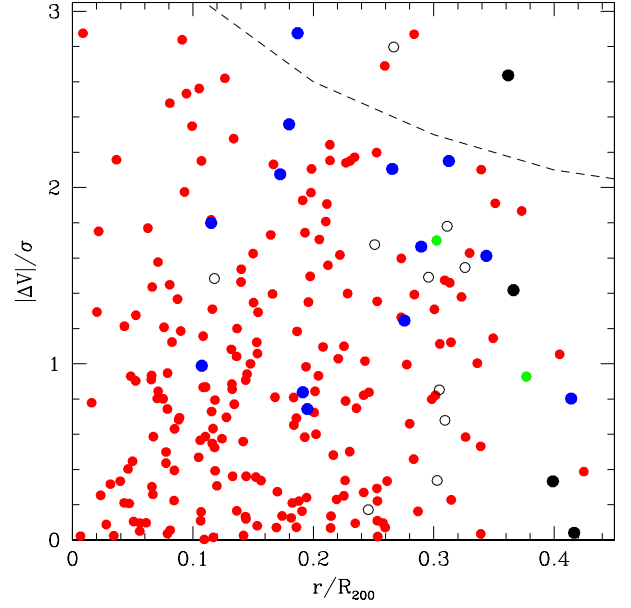}
\caption{Phase-space diagram  of the core of the Coma cluster surveyed in H$\alpha$. 
200 Elliptical+S0 galaxies are indicated in red 
(except two with H$\alpha$ tails in green), 
10 LTGs without  H$\alpha$ extension are given with open symbols, and
17 LTGs with H$\alpha$ tails are over plotted in blue, except NGC 4848 and the 3 galaxies analyzed in this paper  shown
with  filled black symbols.
The dashed line represents the escape velocity in a Coma-like cluster. 
 We adopt for Coma $R_{200}=2.19$ Mpc and $\sigma=880$ $\rm km~s^{-1}$ (Boselli \& Gavazzi 2006). }
\label{phase} 
\end{figure}

\begin{figure}
\centering
 \includegraphics[width=8.5 cm]{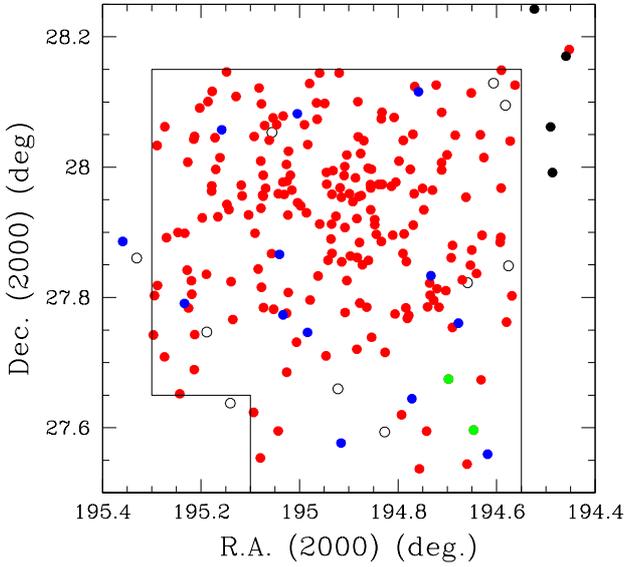}
\caption{Sky distribution of the  core of the Coma cluster surveyed in H$\alpha$ by Yagi et al. (2010) (region within black lines), 
plus the 4 
galaxies observed in this work or by Fossati et al. (2012) (in the NW shown in filled black) and NGC4921 (Kenney et al. 2015 in the far E).
 Same symbols as in Figure \ref{phase}. }
\label{plot} 
\end{figure}

\begin{table*}
\caption{Properties of the  27 surveyed LTG galaxies (plus 2 ETG with tails).}                                                                   
\centering                                                                                        
\begin{tabular}{l c c c c c c c c c c c}                                                                    
\hline\hline                                                                                      

   Jname             & GMP  & CGCG      & NGC   & Other       &   $cz$           & Type  &     $r$   & $g-i$ &Log$M_*$ & $\rm Def_{HI}$ & Ref H$\alpha$  \\
                     &      &           &       &             & $\rm km~s^{-1}$  &       &    mag    &  mag  &$\rm M_{\odot}$   &   &              \\    
 \hline\hline   
SDSSJ125750.2+281013 &      &           &       &             & 6935             &  BCD  &     17.14 &  0.20 &   8.47 & -     & T.W.        \\
SDSSJ125756.7+275930 &      &           &       &             & 4579             &  PEC  &     16.94 &  0.18 &   8.53 & 0.27  & T.W.        \\ 
SDSSJ125757.7+280342 &      &           &       &             & 8147             &  PEC  &     15.24 &  0.70 &   9.75 & 0.53  & T.W.        \\ 
SDSSJ125805.6+281433 &      &   160055  & 4848  &             & 7193             &  Sab  &     13.83 &  0.88 &  10.54 & 0.25  & Fossati+12 \\  
SDSSJ125818.2+275054 &      &           &       &             & 7649             &  BCD  &     17.05 &  0.41 &   8.69 & -     &             \\
SDSSJ125819.7+280541 &      &           &       &             & 7197             &  Sm   &     17.19 &  0.41 &   8.58 & -     &             \\
SDSSJ125825.5+280744 &      &           &       &             & 8212             &  PEC  &     16.23 &  0.62 &   9.23 & -     &             \\
SDSSJ125830.7+273352 &4232  &           &       &             & 7285             &  S..  &     17.75 &  0.41 &   7.39 & -     & Yagi+10    \\    
SDSSJ125838.1+274921 &      &           &       &             & 5424             &  Sm   &     17.35 &  0.56 &   8.67 & -     &             \\
SDSSJ125842.5+274537 &4060  &           &       &             & 8753             &  Sm   &     16.78 &  0.40 &   8.80 & -     & Yagi+10    \\
SDSSJ125856.0+275000 &3896  &   160212  &       &  IC3949     & 7553             &  Sa   &     14.20 &  0.71 &  10.27 & 1.01  & Yagi+10    \\  
SDSSJ125902.0+280656 &3816  &   160213  & 4858  &             & 9430             &  Sb   &     15.21 &  0.62 &   9.70 & 0.96  & Yagi+10    \\ 
SDSSJ125905.2+273840 &3779  &   160073  &       &             & 5435             &  Sa   &     14.37 &  0.56 &  10.00 & 0.82  & Yagi+10    \\ 
SDSSJ125914.9+281503 &      &           &       &             & 7499             &  Sm   &     16.09 &  0.36 &   9.11 & -     &             \\
SDSSJ125939.8+273435 &3271  &           &       &             & 5007             &  PEC  &     15.75 &  0.50 &   9.34 & -     & Yagi+10    \\
SDSSJ125941.3+273935 &      &           &       &             & 6748             &  Sa   &     15.10 &  1.07 &  10.07 & -     &             \\
SDSSJ125956.1+274446 &3071  &           &       &             & 8975             &  BCD  &     16.21 &  0.37 &   9.05 & -     & Yagi+10    \\
SDSSJ130001.0+280455 &3016  &           &       &             & 7770             &  S..  &     17.8  &  0.48 &   7.49 & -     & Yagi+10    \\  
SDSSJ130008.0+274623 &2923  &           &       &             & 8726             &  BCD  &     16.89 &  0.37 &   8.73 & -     & Yagi+10    \\
SDSSJ130009.7+275158 &2910  &   160243  &       &    D100     & 5316             &  BCD  &     15.47 &  0.61 &   9.56 & -     & Yagi+10    \\
SDSSJ130013.4+280311 &      &           &       &             & 8205             &  S... &     17.61 &  0.69 &   8.65 & -     &             \\
SDSSJ130033.7+273815 &      &   160086  &       &             & 7497             &  Sb   &     15.30 &  0.39 &   9.47 & 0.59  &             \\  
SDSSJ130037.9+280326 &2559  &   160252  &       &    IC4040   & 7637             &  Sdm  &     15.07 &  0.60 &   9.78 & 0.45  & Yagi+10    \\ 
SDSSJ130045.2+274449 &      &           &       &             & 9361             &  Sd   &     17.55 &  0.66 &   8.64 & -     &             \\
SDSSJ130056.0+274726 &2374  &   160260  & 4911  &             & 7995             &  Sa   &     13.03 &  1.18 &  11.11 & 0.31  & Yagi+10    \\ 
SDSSJ130119.3+275137 &      &           &       &             & 8260             &  BCD  &     17.02 &  0.63 &   8.86 & -     &             \\
SDSSJ130126.1+275309 &      &   160095  & 4921  &             & 5481             &  Sb   &     12.90 &  1.22 &  11.20 & 0.73  & Kenney+15  \\ 
 \hline
SDSSJ125847.4+274029 &4017  &   160070  & 4854  &             & 8400             &  S0   &     13.86 &  1.22 &  10.74 & -     & Yagi+10     \\
SDSSJ125835.2+273547 &4156  &   160068  & 4853  &             & 7710             &  S0   &     13.51 &  1.03 &  10.78 &  -     & Yagi+10     \\
                                                                                                              
\hline 
\end{tabular}                                                                                     
\label{tabLTG}                                                                                      
\end{table*}    

\begin{figure*}
\centering
\includegraphics[width=19.cm]{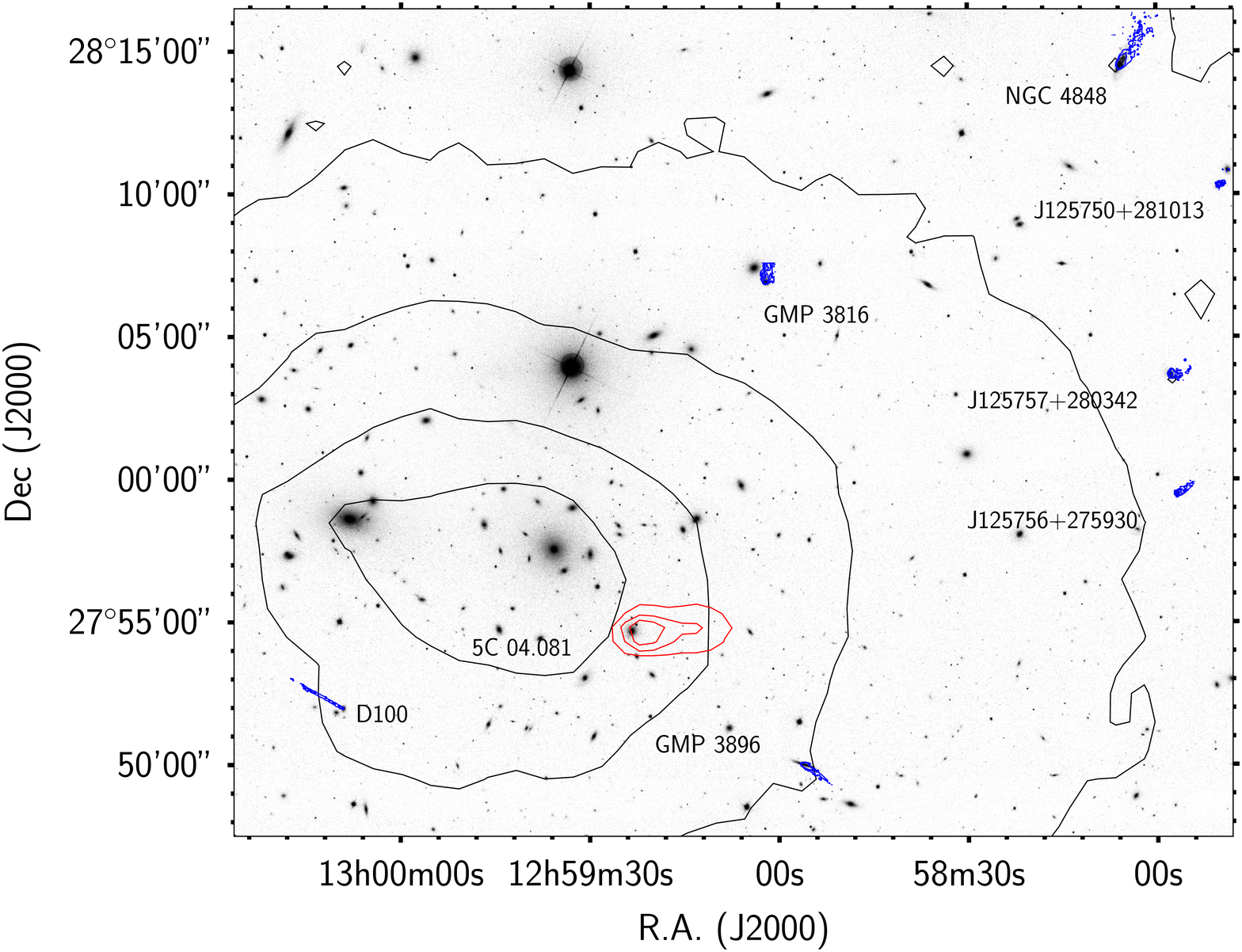}
\caption{Grayscale representation of the distribution of galaxies in the NW part of the Coma cluster (from SDSS)
to highlight the position of the newly discovered H$\alpha$ tails. Black contours represent 
the X-ray emission (XMM 0.4 - 1.3 keV, contour levels: 10, 20, 50, 100, 150 cnts per second per square degree 
with a smoothing of 10 pixel). Red contours represent the radio emission from the head-tail radio source 5C 04.081. Blue contours 
show the H$\alpha$ NET emission  for NGC 4848 from this work, Yagi et al. (2010), and Fossati et al. (2012).
}
\label{coma} 
\end{figure*}
                                                                                                               
\begin{acknowledgements}
This research has made use of the GOLDmine database (Gavazzi et
al. 2003, 2014b) and of the NASA/IPAC Extragalactic Database (NED), which is
operated by the Jet Propulsion Laboratory, California Institute of Technology, under contract with the National 
Aeronautics and Space Administration.  We thank the anonymous referee for his/her constructive criticism.
\end{acknowledgements}

\end{document}